\documentclass[aps,prl,twocolumn,floatfix,superscriptaddress,preprintnumbers,nofootinbib,10pt]{revtex4-1}
\usepackage{geometry}                
\usepackage{graphicx}
\usepackage{amssymb}
\usepackage{amsmath,amssymb}
\usepackage{color}

\definecolor{darkblue}{rgb}{0,0,0.5}
\usepackage[colorlinks,linkcolor=darkblue,citecolor=darkblue,urlcolor=darkblue]{hyperref}

\allowdisplaybreaks

\begin{document}
%%%%%%%%%%%%%%%%%%%%%%%%%%%%%%%%%%%%%%%%%%%%%%%%%%%%%%%%%%%%%%%%%%
\preprint{FERMILAB-PUB-11-509-T}
\preprint{EFI-11-29}

%%%%%%%%%%%%%%%%%%%%%%%%%%%%%%%%%%%%%%%%%%%%%%%%%%%%%%%%%%%%%%%%%%
\title{\boldmath
B Meson Mixing in Effective Theories of \texorpdfstring{\\}{}
Supersymmetric Higgs Bosons}
%%%%%%%%%%%%%%%%%%%%%%%%%%%%%%%%%%%%%%%%%%%%%%%%%%%%%%%%%%%%%%%%%%

\author{Wolfgang Altmannshofer}
\affiliation{Fermi National Accelerator Laboratory, P.O. Box 500, Batavia, IL 60510, USA}

\author{Marcela Carena}
\affiliation{Fermi National Accelerator Laboratory, P.O. Box 500, Batavia, IL 60510, USA}
\affiliation{Enrico Fermi Institute, University of Chicago, Chicago, IL 60637, USA}
\affiliation{Kavli Institute for Cosmological Physics, University of Chicago, Chicago, IL 60637, USA}

%%%%%%%%%%%%%%%%%%%%%%%%%%%%%%%%%%%%%%%%%%%%%%%%%%%%%%%%%%%%%%%%%%
\begin{abstract}

We study a minimal flavor violating extension of the MSSM, where higher dimensional operators in the K\"ahler potential induce tree level non-holomorphic Higgs couplings that are controlled by the scale of the physics beyond the MSSM and analyze their possible impact on CP violation in $B_s$ and $B_d$ mixing. We consider results on the time dependent CP asymmetries in $B_s \to \psi\phi$ and $B_s \to \psi f_0$ from LHCb, in $B_s \to \psi\phi$ from CDF and D0 and in $B \to \psi K_S$ from the B factories as well as the measurement of an anomalous like-sign dimuon charge asymmetry at D0. Taking into account the stringent bounds on the branching ratio of the rare $B_s \to \mu^+ \mu^-$ decay, we investigate to which extent the framework allows to address the observed $(2-3)\sigma$ discrepancies in fits of the unitarity triangle. We find that a non-standard $B_d$ mixing phase, that is in agreement with the current bounds on CP violation in $B_s$ mixing, requires the presence of higher dimensional operators both in the K\"ahler potential and the superpotential. The corresponding region of parameter space is characterized by small $\tan\beta \sim 5$, a light Higgs spectrum with masses below $\lesssim 400$~GeV and will be probed by future measurements at LHCb. 

\end{abstract}
%%%%%%%%%%%%%%%%%%%%%%%%%%%%%%%%%%%%%%%%%%%%%%%%%%%%%%%%%%%%%%%%%%

\maketitle

%%%%%%%%%%%%%%%%%%%%%%%%%%%%%%%%%%%%%%%%%%%%%%%%%%%%%%%%%%%%%%%%%%
\section{Introduction}
%%%%%%%%%%%%%%%%%%%%%%%%%%%%%%%%%%%%%%%%%%%%%%%%%%%%%%%%%%%%%%%%%%

Low Energy observables sensitive to CP Violation in $b \to s$ transitions constitute excellent probes of possible new CP violating phases in extensions of the Standard Model (SM). Indeed, as CP Violation in such processes is predicted to be tiny in the SM, evidence for sizable CP violation in $b \to s$ transitions would be a clear hint for the presence of New Physics (NP). Examples of such low energy probes are observables that are sensitive to the $B_s$ mixing phase, as the semi-leptonic asymmetry $a_{\rm sl}^s$ in decays of $B_s$ mesons to ``wrong sign leptons'' or the time dependent CP asymmetries in the $B_s \to \psi\phi$ and $B_s \to \psi f_0$ decays.

In the context of generic two Higgs doublet models with Minimal Flavor Violation (MFV), where the CKM matrix is the only source of flavor violation, a large $B_s$ mixing phase can be realized if additional CP violating phases are allowed~\cite{Buras:2010mh,Buras:2010zm,Trott:2010iz,Pich:2009sp}. Simultaneously, these models can also address tensions in fits of the CKM matrix that seem to indicate a sizable NP contribution to the $B_d$ mixing phase~\cite{Lunghi:2008aa,Lenz:2010gu}.
Possible relations between a non-standard $B_s$ mixing phase and the baryon asymmetry of the universe in these models have been studied in~\cite{Cline:2011mm,Liu:2011jh}.

In the context of the Minimal Supersymmetric Standard Model (MSSM) on the other hand, a MFV soft sector is not sufficient to generate sizable NP phases in meson mixing~\cite{Altmannshofer:2008hc,Altmannshofer:2009ne} due to the strong experimental bound on the $B_s \to \mu^+ \mu^-$ branching ratio.
Non SM-like $B_s$ and $B_d$ mixing phases in the MSSM require new sources of flavor violation in addition to the CKM matrix (see e.g.~\cite{Altmannshofer:2009ne,Ko:2008xb} for studies of such frameworks).
Supersymmetric models with MFV {\it can} generate large $B$ mixing phases {\it if} they allow for a strongly reduced muon Yukawa coupling such that the BR$(B_s \to \mu^+ \mu^-)$ constraint can be avoided. Such a situation can be realized for example in the so-called uplifted SUSY Higgs region~\cite{Dobrescu:2010mk}, even though this framework is strongly constrained by other $B$ physics observables and $(g-2)_\mu$~\cite{Altmannshofer:2010zt}.
As studied in detail in~\cite{Altmannshofer:2011rm}, non-negligible corrections to CP violating observables in meson mixing can also be generated if the MSSM with MFV is extended by the two leading higher dimensional operators in the Higgs sector with complex coefficients~\cite{Dine:2007xi}. Still, the stringent bounds on BR$(B_s \to \mu^+ \mu^-)$ only allow for a $B_s$ mixing phase of $S_{\psi\phi} \lesssim 0.15$ in specific regions of parameter space of this model.

In this work we analyze an extension of the MSSM, introducing higher dimensional operators not exclusively in the Higgs sector, but also considering dimension 5 operators that induce non-holomorphic Higgs-fermion couplings already at the tree level~\cite{Antoniadis:2008es}. Such operators are a possible source of flavor and CP violation.
Assuming that these operators obey the Minimal Flavor Violation ansatz~\cite{Chivukula:1987py,Buras:2000dm,D'Ambrosio:2002ex}, we explore to which extent the considered framework allows for non-standard CP violation in $B$ mixing without being in conflict with the bounds on the BR$(B_s \to \mu^+\mu^-)$. 
For a study of a similar framework, that however does not consider CP violation in $B$ mixing, see~\cite{Bae:2010ai}.

%%%%%%%%%%%%%%%%%%%%%%%%%%%%%%%%%%%%%%%%%%%%%%%%%%%%%%%%%%%%%%%%%%
\section{\boldmath CP Violation in \texorpdfstring{$B$}{B} Meson Mixing}
%%%%%%%%%%%%%%%%%%%%%%%%%%%%%%%%%%%%%%%%%%%%%%%%%%%%%%%%%%%%%%%%%%

The $B_q$ mixing amplitude
\begin{equation}
\mathcal{M}^q = M^q_{12} - \frac{i}{2}\Gamma^q_{12}~,
\end{equation}
consists on an dispersive part, $M_{12}^q$, and an absorptive part, $\Gamma_{12}^q$.
The absorptive part is dominated by tree level SM contributions and therefore hardly affected in many NP models. Throughout this work we will assume that $\Gamma_{12}^q$ has no significant NP contributions. The dispersive part $M_{12}^q$ on the other hand is highly sensitive to new heavy degrees of freedom. The effects of NP in $M_{12}^q$ can be parametrized by
\begin{equation}
M_{12}^q = C_q e^{i \phi_q^{\rm NP}} (M_{12}^q)_{\rm SM} ~.
\end{equation}
The main impact of the parameters $C_d$ and $C_s$ is on the mass differences $\Delta M_d$ and $\Delta M_s$, respectively.
The NP phases $\phi_d^{\rm NP}$ and $\phi_s^{\rm NP}$ affect observables that are sensitive to CP violation in B meson mixing, like the semileptonic asymmetries $a_{\rm SL}^d$ and $a_{\rm SL}^s$ as well as the time-dependent CP asymmetries in $B_d \to \psi K_S$, $B_s \to \psi\phi$ and $B_s \to \psi f_0$.
In order to constrain the NP parameters through measurements of these observables, knowledge of the respective SM contributions is required. As several of the observables, in particular $\Delta M_d$ and $S_{\psi K_S}$ -- the time dependent CP asymmetry in $B_d \to \psi K_S$ -- are a key ingredient in the determination of the Unitarity Triangle (UT), a simultaneous fit of the CKM parameters and the NP parameters as performed in~\cite{Lenz:2010gu} is the most consistent approach. Here we focus mainly on the impact of the recent improvements on the determination of the $B_s$ mixing phase at CDF and D0~\cite{Aaltonen:2007he,CDFD0update} and in particular at LHCb~\cite{LHCb}. 
As these measurements have no significant effect on the determination of the CKM parameters and the NP parameters other than $\phi_s^{\rm NP}$, we consider a simplified approach and take the CKM parameters as well as $C_d$ and $C_s$ from the generic NP fit in~\cite{Lenz:2010gu}, and fit only the NP parameters $\phi_s^{\rm NP}$ and $\phi_d^{\rm NP}$. We expect this approach to give a good estimate of the allowed region of parameter space compatible with the present experimental data on B meson mixing. In particular we will use
\begin{eqnarray} \label{eq:beta}
\beta &=& \textnormal{Arg}\left[ (V_{tb} V_{td}^*) / (V_{cb} V_{cd}^*)\right] = (27.2^{+1.1}_{-3.1})^\circ \\ 
\beta_s &=& \textnormal{Arg}\left[ (V_{tb} V_{ts}^*)/(V_{cb} V_{cs}^*)\right] = (-1.3\pm 0.1)^\circ \nonumber
\end{eqnarray}
and the following $2\sigma$ bounds
\begin{equation}
0.62 < C_d < 1.15 ~~,~~~ 0.79 < C_s < 1.23~. 
\end{equation}

We now give expressions for the observables that are sensitive to CP violation in B meson mixing.
For the semileptonic asymmetries we obtain~\cite{Lenz:2006hd,Lenz:2010gu}
\begin{eqnarray} \label{eq:asld}
10^4\,a_{\rm SL}^d &\simeq& (55.2 \sin\phi_d^{\rm NP} - 4.80 \cos\phi_d^{\rm NP})/C_d ~~ \\ \label{eq:asls}
10^4\,a_{\rm SL}^s &\simeq& (49.7 \sin\phi_s^{\rm NP} + 0.19 \cos\phi_s^{\rm NP})/C_s ~~
\end{eqnarray}
where the uncertainties on the numerical coefficients are at the level of $\sim 15\%$. 
For the time dependent CP asymmetries one has
\begin{eqnarray} \label{eq:SpsiKs}
S_{\psi K_S} &=& \sin(2\beta + \phi_d^{\rm NP}) ~, \\
S_{\psi \phi} = S_{\psi f_0} &=& \sin(2|\beta_s| - \phi_s^{\rm NP}) ~.
\end{eqnarray}
These expressions hold under the usual assumption that the $B_d \to \psi K_S$, $B_s \to \psi\phi$ and $B_s \to \psi f_0$ decays are dominated by the SM tree level amplitudes. 
Using~(\ref{eq:beta}) the corresponding SM predictions are
\begin{eqnarray}
S_{\psi K_S}^{\rm SM} = 0.82^{+0.02}_{-0.07} ~,~ S_{\psi \phi}^{\rm SM} = 0.046^{+0.002}_{-0.003} ~.
\end{eqnarray}

The D0 collaboration measured the like-sign dimuon charge asymmetry that is predicted to be composed out of the semi-leptonic asymmetries in the $B_d$ and $B_s$ decays~\cite{Abazov:2011yk}
\begin{eqnarray}
A_{\rm SL}^b &=& 0.59 a_{\rm SL}^d + 0.41 a_{\rm SL}^s \nonumber \\ \label{eq:aslb_D0}
&=& (-78.7 \pm 19.6)10^{-4} ~.
\end{eqnarray}
The corresponding SM prediction $A_{\rm SL}^b(\rm SM) = (-2.8^{+0.5}_{-0.6})10^{-4}$~\cite{Lenz:2010gu} is roughly a factor 25 below the central value in~(\ref{eq:aslb_D0}) and differs from it by $3.9\sigma$. The value~(\ref{eq:aslb_D0}) updates an earlier D0 study~\cite{Abazov:2010hv} that found a $3.2\sigma$ evidence for an anomalous like-sign dimuon charge asymmetry.
A separate extraction of the semileptonic asymmetries results in~\cite{Abazov:2011yk}
\begin{eqnarray} \label{eq:asld_D0}
a_{\rm SL}^d(D0) &=& (-12 \pm 52)10^{-4} ~,\\ \label{eq:asls_D0}
a_{\rm SL}^s(D0) &=& (-181 \pm 106)10^{-4} ~.
\end{eqnarray}
The results in Eqs.~(\ref{eq:aslb_D0}),~(\ref{eq:asld_D0}) and~(\ref{eq:asls_D0}) hint towards large negative values for the NP phases $\phi_d^{\rm NP}$ and, in particular, $\phi_s^{\rm NP}$. 

Interestingly enough, there is a (2-3)$\sigma$ tension between the SM prediction of $S_{\psi K_S}$~(\ref{eq:SpsiKs}) and its experimental value~\cite{Asner:2010qj}
\begin{equation}
S_{\psi K_S}^{\rm exp} = 0.67 \pm 0.02 ~,
\end{equation}
that is largely driven by the $B \to \tau \nu$ measurements that prefer a large value of $|V_{ub}|$. 
This tension points in the same direction for the NP phase $\phi_d^{\rm NP}$ as the data on the like sign dimuon charge asymmetry. 

A small preference for a negative NP phase in $B_s$ mixing was also observed in CDF and D0 data on the time dependent CP asymmetry in $B_s \to \psi\phi$ that give @ 95\%~C.L. ~\cite{Aaltonen:2007he,CDFD0update}
\begin{eqnarray}
-1.36 < &\phi_s^{\rm NP} - 2|\beta_s|& < 0.26 ~~{\rm (CDF)}~, \\
-1.65 < &\phi_s^{\rm NP} - 2|\beta_s|& < 0.24 ~~{\rm (D0)}~.
\end{eqnarray}
Combining the results from~\cite{Aaltonen:2007he} with~\cite{Abazov:2010hv}, global fits to the data found a $B_s$ mixing phase $\phi_s^{\rm NP} = O(-1)$~\cite{Ligeti:2010ia,Lenz:2010gu}.

%%%%%%%%%%%%%%%%%%%%%%%%%%%%%%%%%%%%%%%%%%%%%%%%%%%%%%%%%%%%%%%
\begin{figure}[t] \centering
\includegraphics[width=0.47\textwidth]{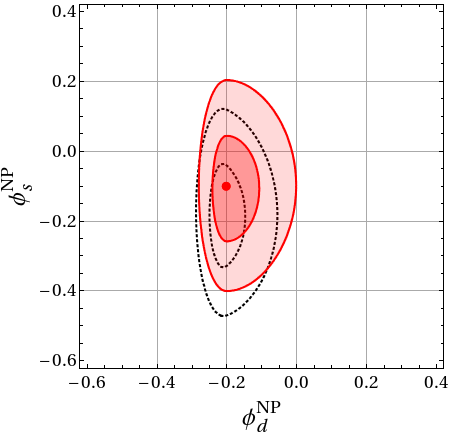}
\caption{
Allowed ranges for the NP phases $\phi_s^{\rm NP}$ and $\phi_d^{\rm NP}$ at the 1 and 2 $\sigma$ level, taking into account the measurements of the time-dependent CP asymmetries in $B_s \to \psi\phi$ at CDF, D0 and LHCb~\cite{Aaltonen:2007he,LHCb}, in $B_s \to \psi f_0$ at LHCb~\cite{LHCb} and in $B_d \to \psi K_S$ at the B factories~\cite{Asner:2010qj}. The measurement of the like-sign dimuon charge asymmetry at D0~\cite{Abazov:2011yk} is included in the black dotted contours but {\it not} in the red solid contours.
}
\label{fig:phi_fit}
\end{figure}
%%%%%%%%%%%%%%%%%%%%%%%%%%%%%%%%%%%%%%%%%%%%%%%%%%%%%%%%%%%%%%%

Recently however, LHCb presented results on the time dependent CP asymmetries in $B_s \to \psi\phi$ and $B_s \to \psi f_0$ that are consistent with the tiny SM prediction and that strongly restrict the possible values for a NP phase in $B_s$ mixing~\cite{LHCb}
\begin{equation}
\phi_s^{\rm NP} - 2|\beta_s| = 0.03 \pm 0.16 \pm 0.07 ~.
\end{equation}
In Fig.~\ref{fig:phi_fit} we show the result of a simple fit of the NP phases $\phi_d^{\rm NP}$ and $\phi_s^{\rm NP}$ to the combined LHCb result on the time-dependent CP asymmetries in $B_s \to \psi\phi$ and $B_s \to \psi f_0$~\cite{LHCb}, the results on the time-dependent CP asymmetry in $B_s \to \psi\phi$ from CDF and D0~\cite{Aaltonen:2007he,CDFD0update} as well as the measurement of $S_{\psi K_S}$ at the B factories~\cite{Asner:2010qj}, using the values for $\beta$ and $\beta_s$, $C_d$ and $C_s$ from above.
The allowed region is mainly determined by the measurements of $S_{\psi\phi}$, $S_{\psi f_0}$ at LHCb and $S_{\psi K_S}$ at the B factories, while the measurements of $S_{\psi\phi}$ at CDF and D0 lead to a small shift of the central value of $\phi_s^{\rm NP}$ towards a small negative value. We stress that the very large value of the like-sign dimuon charge asymmetry observed by D0 cannot be explained given the current data on the time dependent CP asymmetries in $B_s \to \psi\phi$ and $B_s \to \psi f_0$. Using our fit results we find a central value of $A_{\rm SL}^b = -11 \cdot 10^{-4}$ and a $2\sigma$ range of $-18 < 10^4\, A_{\rm SL}^b  < -2$. This differs from the measured value~(\ref{eq:aslb_D0}) by $\simeq 3 \sigma$.
Including the like-sign dimuon charge asymmetry directly into the fit leads only to a small shift towards slightly larger negative $\phi_s^{\rm NP}$ values. In the following we focus on the fit that does not include the $A_{\rm sl}^b$ measurement.

Due to the small discrepancy between the experimental determination of $S_{\psi K_S}$ and its SM prediction coming from the UT fits~\cite{Lunghi:2008aa,Lenz:2010gu}, the NP phase in $B_d$ mixing shows preference towards a negative value
\begin{equation} \label{eq:phi_d} 
\phi_d^{\rm NP} = -0.20^{+0.10}_{-0.04}~,
\end{equation}
that is roughly 2$\sigma$ below 0.
While also for the NP phase in $B_s$ mixing we find a slight preference for a small negative value, $\phi_s^{\rm NP}$ is perfectly consistent with zero
\begin{equation} \label{eq:phi_s}
\phi_s^{\rm NP} = -0.10 \pm 0.15 ~.
\end{equation}
Presently, this still leaves room for NP contributions, but the bound~(\ref{eq:phi_s}) will improve significantly in the near future with more data from LHCb.

%%%%%%%%%%%%%%%%%%%%%%%%%%%%%%%%%%%%%%%%%%%%%%%%%%%%%%%%%%%%%%%%%%
\section{\boldmath The \texorpdfstring{$B_q \to \mu^+\mu^-$}{Bq --> mu+mu-} Decays}
%%%%%%%%%%%%%%%%%%%%%%%%%%%%%%%%%%%%%%%%%%%%%%%%%%%%%%%%%%%%%%%%%%

The CDF collaboration reported a small excess in $B_s \to \mu^+\mu^-$ candidates~\cite{Aaltonen:2011fi}, leading to
\begin{equation} \label{eq:Bsmm_CDF}
{\rm BR}(B_s \to \mu^+ \mu^-)_{\rm CDF} = (1.8^{+1.1}_{-0.9}) \times 10^{-8}~, 
\end{equation}
No excess has been observed by LHCb and CMS that report the following bounds~\cite{Chatrchyan:2011kr} 
\begin{eqnarray} \label{eq:Bsmm_LHCb}
{\rm BR}(B_s \to \mu^+ \mu^-)_{\rm LHCb} & < &   1.5 \cdot  10^{-8} ~, \\ \label{eq:Bsmm_CMS}
{\rm BR}(B_s \to \mu^+ \mu^-)_{\rm CMS~} & < &   1.9 \cdot  10^{-8} ~.
\end{eqnarray}
Combining the bounds from LHCb and CMS one finds~\cite{LHCbCMS}
\begin{equation} \label{eq:Bsmm_LHC}
{\rm BR}(B_s \to \mu^+ \mu^-)_{\rm LHC} < 1.1 \times 10^{-8} 
\end{equation}
that is only a factor of 3.5 above the SM prediction~\cite{Buras:2003td}
\begin{equation}
{\rm BR}(B_s \to \mu^+ \mu^-)_{\rm SM} = (3.2 \pm 0.2) \times 10^{-9}~. 
\end{equation}
The current bounds on the $B_d \to \mu^+ \mu^-$ branching ratio~\cite{Chatrchyan:2011kr} are still a factor 40-50 above the SM expectation and therefore $B_d \to \mu^+ \mu^-$ is much less constraining than $B_s \to \mu^+ \mu^-$ in models with MFV.

Given the strong bound on BR$(B_s \to \mu^+ \mu^-)$, possible neutral Higgs contributions to B mixing in the MSSM with MFV are strongly constrained~\cite{Buras:2002vd,Carena:2006ai}. Also in the BMSSM model considered in~\cite{Altmannshofer:2011rm} non-standard B mixing phases are rather restricted ($S_{\psi\phi} \lesssim 0.15$) and can only be generated in particular corners of parameter space.

In the following we present an extension of the MSSM that respects the MFV principle but allows nonetheless for sizable NP phases in B mixing without being in conflict with the BR$(B_s \to \mu^+ \mu^-)$ constraint.

%%%%%%%%%%%%%%%%%%%%%%%%%%%%%%%%%%%%%%%%%%%%%%%%%%%%%%%%%%%%%%%%%%
\section{Effective Higgs - Fermion Couplings Beyond the MSSM}
%%%%%%%%%%%%%%%%%%%%%%%%%%%%%%%%%%%%%%%%%%%%%%%%%%%%%%%%%%%%%%%%%%

At the tree level, the MSSM is a 2 Higgs doublet model of type II and the couplings of the neutral Higgs bosons to fermions are flavor conserving. At the loop level on the other hand, non-holomorphic Higgs couplings are generated and have important consequences. Loop induced couplings of down-type quarks and charged leptons to the up-type Higgs can lead to large threshold corrections to the corresponding masses~\cite{Hempfling:1993kv} and modify significantly CKM matrix elements~\cite{Blazek:1995nv} as well as charged Higgs couplings to quarks~\cite{Carena:1999py}. Finally, they also generate flavor changing neutral Higgs couplings that can have a profound impact on flavor phenomenology~\cite{Hamzaoui:1998nu,Buras:2002vd,Carena:2006ai}.
All these effects become relevant for large values of $\tan\beta$ that can compensate for the 1-loop suppression.

We now consider possible extensions of the MSSM with new degrees of freedom at a scale $M \simeq $~several~TeV. As long as the SUSY breaking scale $m_S$ of the new degrees of freedom is small compared to M and as long as M is sufficiently larger than the scale of the MSSM degrees of freedom, one can describe the effects of the Beyond the MSSM (BMSSM) physics by higher dimensional operators suppressed by $1/M$~\cite{Brignole:2003cm,Dine:2007xi}. 
An analysis up to order $1/M$ captures the physics of several MSSM UV extensions while the effective description of others need to include $1/M^2$ effects~\cite{Carena:2009gx,Antoniadis:2009rn}. In this work we restrict ourselves to the $1/M$ level. 
We consider both the leading higher dimensional superpotential operators that involve only Higgs fields~\cite{Dine:2007xi}
\begin{equation} \label{eq:super}
\mathcal{L} \supset \frac{\omega}{2 M} \int d^2\theta (1 + \alpha Z) (H_u H_d)^2 
\end{equation}
and in particular also $1/M$ suppressed K\"ahler potential operators that induce non-holomorphic Higgs-fermion couplings already at the tree level~\cite{Dine:2007xi,Antoniadis:2008es} 
\footnote{Possible UV completions that lead to the operators in~(\ref{eq:super}) and~(\ref{eq:kahler}) are discussed e.g. in~\cite{Dine:2007xi,Antoniadis:2008es,Carena:2009gx,Bae:2010ai} and briefly presented in the Appendix.}
\begin{eqnarray} \label{eq:kahler}
\mathcal{L} && \supset \frac{1}{M} \int d^4\theta \left( 1 + Z + Z^\dagger + Z Z^\dagger \right) \\
\times && \left( \lambda_u H_d^\dagger Q U + \lambda_d  H_u^\dagger Q D + \lambda_\ell H_u^\dagger L E\right) + {\rm h.c.} ~. \nonumber
\end{eqnarray}
In the above expressions, $Z$ is an auxiliary dimensionless spurion that develops a SUSY breaking F-term
\begin{equation}
Z \to m_S \theta^2~.
\end{equation}
The phenomenological consequences of the operators in~(\ref{eq:super}) have been thoroughly studied in the literature~\cite{Dine:2007xi,Carena:2009gx,Antoniadis:2007xc,Blum:2010by,Altmannshofer:2011rm}. In particular, they can significantly enhance the tree level mass of the lightest Higgs boson of the MSSM and lead to sizable mass splittings between the two heavy neutral Higgs bosons and also the charged Higgs boson. Possible phases of the coefficients $\alpha$ and $\omega$ lead also to scalar-pseudoscalar mixing. Their impact in the context of Higgs and flavor phenomenology has been analyzed in~\cite{Altmannshofer:2011rm}.

The K\"ahler potential operators~(\ref{eq:kahler}) modify the interactions of Higgs bosons, (s)quarks and (s)leptons of the MSSM at the $1/M$ level~\cite{Dine:2007xi,Antoniadis:2008es,Bae:2010ai}. 
The supersymmetric part of~(\ref{eq:kahler}) for example leads to corrections of the holomorphic MSSM Yukawa couplings
\begin{equation}
Y_f \to Y_f^\prime = Y_f + \frac{\mu}{M} \lambda_f ~. 
\end{equation}
After SUSY breaking also non-holomorphic Higgs-quark couplings are generated
\begin{eqnarray}
\mathcal{L} &\supset& \frac{m_S}{M} (\lambda_u)_{ij} ~H_d^\dagger \bar Q_i U_j + \frac{m_S}{M} (\lambda_d)_{ij} ~H_u^\dagger \bar Q_i D_j \nonumber \\
&& + \frac{m_S}{M} (\lambda_\ell)_{ij} ~H_u^\dagger \bar L_i E_j  ~+ {\rm h.c.}~,
\end{eqnarray}
where we now made flavor indices explicit. These terms can lead to flavor changing neutral Higgs vertices and correspondingly to tree level contributions to FCNC processes like $B$ mixing and $B_s \to \mu^+\mu^-$. In the following we will focus on them.
The MFV hypothesis as formulated in~\cite{D'Ambrosio:2002ex} amounts to the assumption that the $SU(3)^3$ quark flavor symmetry of the gauge sector is broken by only two spurions $\mathcal{Y}_u$ and $\mathcal{Y}_d$ that transform as $(3,\bar 3,1)$ and $(3,1,\bar 3)$ respectively. Correspondingly, the couplings $Y_q$ and $\lambda_q$ can be expanded in powers of these spurions. To keep notation simple and concise, we conveniently choose  $\mathcal{Y}_q = Y_q^\prime = Y_q + \frac{\mu}{M} \lambda_q$ to be these spurions. Any other linear combination of $Y_q$ and $\lambda_q$ leads to equivalent results. For the non-holomorphic Higgs couplings $\lambda_q$ to the down quarks one then has
\begin{eqnarray} \label{eq:non_holomorphic}
\lambda_d &=& \varepsilon_0 Y_d + \varepsilon_1 Y_d Y_d^\dagger Y_d + \varepsilon_2 Y_u Y_u^\dagger Y_d + \\
&+& \varepsilon_3 Y_u Y_u^\dagger Y_d Y_d^\dagger Y_d + \varepsilon_4 Y_d Y_d^\dagger Y_u Y_u^\dagger Y_d + \dots ~, \nonumber
\end{eqnarray}
where for simplicity we dropped the prime on the corrected Yukawa couplings.
An analogous expression holds for the up quark coupling which is however not relevant for the following discussion.
The coefficients $\varepsilon_i$ are generically of O(1) and complex.
For later convenience we define
\begin{eqnarray}
\bar\varepsilon_0 &=& \frac{m_S}{M} \varepsilon_0 ~~, \nonumber \\
\bar\varepsilon_1 &=& \frac{m_S}{M} y_b^2 \varepsilon_1 ~~,~~~ \bar\varepsilon_3 = \frac{m_S}{M} y_t^2 y_b^2 \varepsilon_3 ~,\nonumber \\
\bar\varepsilon_2 &=& \frac{m_S}{M} y_t^2 \varepsilon_2 ~~,~~~ \bar\varepsilon_4 = \frac{m_S}{M} y_t^2 y_b^2 \varepsilon_4 ~,\nonumber
\end{eqnarray}
\begin{eqnarray}
\bar\varepsilon_5 &=& \bar\varepsilon_0 + \bar\varepsilon_1 + \bar\varepsilon_2 + \bar\varepsilon_3 + \bar\varepsilon_4~, \nonumber \\
\bar\varepsilon_6 &=& \bar\varepsilon_0 + \bar\varepsilon_1 + \bar\varepsilon_4 ~.
\end{eqnarray}
The couplings in~(\ref{eq:non_holomorphic}) modify the relation between the down quark masses $m_q$ and the corresponding Yukawa couplings $y_q$
\begin{eqnarray} \label{eq:masses}
\frac{y_d v}{m_d} = \frac{y_s v}{m_s} &=& \frac{\tan\beta}{1+\bar\varepsilon_0 \tan\beta} \nonumber \\
\frac{y_b v}{m_b} &=& \frac{\tan\beta}{1+ \bar\varepsilon_5 \tan\beta}~,
\end{eqnarray}
with the Higgs vev $v = 174$~GeV and we only show the leading term in a $\tan\beta$ expansion. 

Similar to the quark masses, also the CKM matrix receives $\tan\beta$ enhanced corrections. The relations between the affected elements of the bare CKM matrix $V^0$ and the physical CKM matrix $V$ read ($i = 1,2$)

\begin{equation} \label{eq:CKM}
\frac{V_{ti}^0}{V_{ti}} = \left(\frac{V_{ib}^0}{V_{ib}}\right)^* = \frac{1 + \bar\varepsilon_5^* \tan\beta}{1 + \bar\varepsilon_6^* \tan\beta} ~.
\end{equation}
Finally we also give explicit expressions for the corrected flavor changing couplings of right handed down quarks with the Higgs bosons. The leading $\tan\beta$ enhanced terms read ($i \neq j$)
\begin{widetext}
\begin{equation} \label{eq:FCNC_Lagrangian}
\mathcal{L} \supset \bar d_L^i \frac{m_{d_j}}{v} V_{ti}^*V_{tj} ~X_{ij}~ d_R^j \left( c_\alpha H - s_\alpha h + i A\right) ~+~ \bar u_L^i \frac{m_{d_j}}{v} V_{ij} ~Z_{ij}~ d_R^j H^+  ~+ \textnormal{h.c.}
\end{equation}
\begin{eqnarray} \label{eq:Higgs_couplings}
Z_{ib} &=& \frac{t_\beta}{1+\bar\varepsilon_6 t_\beta} ~,~~~~~~~~ X_{ib} = - \frac{(\bar\varepsilon_2 + \bar\varepsilon_3)~t^2_\beta}{(1 + \bar\varepsilon_5 t_\beta)(1 + \bar\varepsilon_6 t_\beta)}~, \\
X_{bi} &=& - \frac{(\bar\varepsilon_2 + \bar\varepsilon_4)~t^2_\beta}{(1 + \bar\varepsilon_5 t_\beta)(1 + \bar\varepsilon_6 t_\beta)}~ \left[ \frac{1+\bar\varepsilon_6 t_\beta}{1+\bar\varepsilon_0 t_\beta} - \frac{1+\bar\varepsilon_6 t_\beta}{1+\bar\varepsilon_6^* t_\beta}~ \frac{\bar\varepsilon_2^* + \bar\varepsilon_3^*}{\bar\varepsilon_2 + \bar\varepsilon_4}~ \frac{(\bar\varepsilon_1 + \bar\varepsilon_3)t_\beta}{1+\bar\varepsilon_0 t_\beta} \right]~. \label{eq:Xbi}
\end{eqnarray}
\end{widetext}
The flavor changing $b_R \to d_L^i$ couplings $X_{ib}$ are generated by the $\varepsilon_2$ and $\varepsilon_3$ terms and they are proportional to $m_b$. The flavor changing $b_L \to d_R^i$ couplings $X_{bi}$ on the other hand are generated by the $\varepsilon_2$ and $\varepsilon_4$ terms and suppressed by light quark masses. 
The expression~(\ref{eq:Xbi}) generalizes the results given in~\cite{D'Ambrosio:2002ex,Carena:2006ai,Hofer:2009xb} and, to the best of our knowledge, has not been presented in the literature.

In the MSSM, the $\bar\epsilon$ factors can only be loop induced. Gluino-down squark loops generate for example $\bar\epsilon_0$, while $\bar\epsilon_2$ is generated by chargino-stop loops. Due to the loop suppression, the $\bar\epsilon$ factors are generically of O(0.01) in the MSSM. Correspondingly, the corrections in~(\ref{eq:masses}),~(\ref{eq:CKM}),~(\ref{eq:Higgs_couplings}) and~(\ref{eq:Xbi}) become important only for large values of the ratio of the two Higgs vevs $v_u/v_d = \tan\beta = t_\beta$. On the other hand, in generic 2 Higgs doublet models with MFV as analyzed in~\cite{Buras:2010mh,Buras:2010zm}, where the $\bar\epsilon$ are free parameters, moderate values of $\tan\beta \simeq 5 - 10$ are sufficient to generate O(1) effects. The same is true in the supersymmetric framework considered here as long as the BMSSM scale that controls the size of the $\bar\epsilon$ factors is not too high, i.e. $m_S / M \sim 0.1$.  

%%%%%%%%%%%%%%%%%%%%%%%%%%%%%%%%%%%%%%%%%%%%%%%%%%%%%%%%%%%%%%%%%%
\section{Higgs Effects in Flavor Physics} \label{sec:flavor}
%%%%%%%%%%%%%%%%%%%%%%%%%%%%%%%%%%%%%%%%%%%%%%%%%%%%%%%%%%%%%%%%%%

Observables that are highly sensitive to a non-standard Higgs sector are observables in meson mixing as well as the branching ratios of the rare decays $B_{s,d} \to \mu^+\mu^-$ that receive tree level contributions from flavor changing neutral Higgs exchange. 
Charged Higgs effects are relevant in the $B \to X_s \gamma$ decay as well as in the $B \to \tau \nu$, $B \to D \tau \nu$ and $K \to \mu \nu$ decays. The first one is modified only at the loop level while the others receive contributions from tree level charged Higgs exchange. However these tree level decays turn out to give only mild constraints in regions of parameter space with non-standard B meson mixing phases and we do not discuss them in detail here although they are included in our numerical analysis. For a recent study of $B \to \tau \nu$ in the context of multi Higgs doublet models with MFV see~\cite{Blankenburg:2011ca}.

%%%%%%%%%%%%%%%%%%%%%%%%%%%%%%%%%%%%%%%%%%%%%%%%%%%%%%%%%%%%%%%%%%
\subsection{\boldmath \texorpdfstring{$B_s$ - $\bar B_s$}{Bs - Bsbar} Mixing}
%%%%%%%%%%%%%%%%%%%%%%%%%%%%%%%%%%%%%%%%%%%%%%%%%%%%%%%%%%%%%%%%%%

The flavor changing neutral Higgs couplings in~(\ref{eq:FCNC_Lagrangian}) give rise to tree level contributions to $B_s$ mixing mediated by neutral Higgs exchange.
These contributions can be described by the following effective Hamiltonian
\begin{equation}
\mathcal{H}_\textnormal{eff} = \tilde  C_2 (\bar b_R s_L)^2 + C_4 (\bar b_R s_L)(\bar b_L s_R) ~,
\end{equation}
with the Wilson coefficients
\begin{eqnarray}\label{eq:Wilson2}
\tilde C_2 &\simeq& -\frac{1}{2} ~X_{sb}^2 ~(V_{tb}V_{ts}^*)^2~ \frac{m_b^2}{M_A^2} ~ \frac{\alpha \omega m_S}{M} \frac{1}{M_A^2} ~, \\ \label{eq:Wilson4}
C_4 &\simeq& X_{sb} X_{bs}^* ~(V_{tb}V_{ts}^*)^2~ \frac{m_b m_s}{v^2} ~\frac{2}{M_A^2} ~.
\end{eqnarray}
Analogous contributions to $B_d$ mixing can be obtained through the replacements $s \to d$.\footnote{Neutral Higgs contributions to Kaon mixing are proportional to $m_s^2$ or $m_s m_d$ in the considered framework and therefore negligibly small.}
In writing~(\ref{eq:Wilson2}) and~(\ref{eq:Wilson4}), we assume the decoupling limit $M_A \gg M_Z$ and treat the effect of the operators in~(\ref{eq:super}) in a mass insertion approximation. In our numerical analysis instead we work with Higgs mass eigenstates that we derive from the full Higgs potential including MSSM 2-loop corrections~\cite{Carena:1995bx}.

The Wilson coefficient $C_4$ is proportional to $m_b m_q$. Consequently it can only lead to sizable effects in $B_s$ mixing while its impact on $B_d$ mixing is rather restricted. On the other hand, $\tilde C_2$ is proportional to $m_b^2$ and therefore leads to NP contributions of the same size and phase both in $B_s$ and $B_d$ mixing.
We stress that $C_4$ is complex only if higher orders of the bottom Yukawa are considered in the expansion~(\ref{eq:non_holomorphic}) (see also~\cite{Kagan:2009bn}). Indeed, one easily checks that switching off $\bar\varepsilon_1$, $\bar\varepsilon_3$ and $\bar\varepsilon_4$ leads to a real $C_4$.
The Wilson coefficient $\tilde C_2$ however is also highly sensitive to the phases of $\bar\varepsilon_0$, $\bar\varepsilon_2$, $\omega$ and $\alpha$.
From~(\ref{eq:Wilson2}) it is clear that $\tilde C_2$ is only relevant in presence of the higher dimensional operators~(\ref{eq:super}) in the superpotential and for small Higgs masses not far above the electroweak scale. 

%%%%%%%%%%%%%%%%%%%%%%%%%%%%%%%%%%%%%%%%%%%%%%%%%%%%%%%%%%%%%%%%%%
\subsection{\boldmath The \texorpdfstring{$B_s \to \mu^+\mu^-$}{Bs --> mu+mu-} Decay}
%%%%%%%%%%%%%%%%%%%%%%%%%%%%%%%%%%%%%%%%%%%%%%%%%%%%%%%%%%%%%%%%%%

The $B_s \to \mu^+ \mu^-$ decay receives tree level contributions from flavor changing neutral Higgs exchange.
One finds 
\begin{eqnarray}
R_{B_s\mu\mu} &=& \frac{{\rm BR}(B_s \to \mu^+ \mu^-)_{\phantom{\rm SM}}}{{\rm BR}(B_s \to \mu^+ \mu^-)_{\rm SM}} \nonumber \\
&\simeq& \left|A\right|^2 + \left|1- A\right|^2 ~,
\end{eqnarray}
\begin{equation} \label{eq:A}
A \simeq - \left(\frac{4\pi}{\alpha_2}\right) \frac{X_{sb}}{4 Y_0(x_t)} \frac{t_\beta}{(1+\bar\epsilon_\ell^* t_\beta)} \frac{m_{B_s}^2}{M_A^2} ~,
\end{equation}
with the SM loop function given by $Y_0(x_t) \simeq 0.96$.
The above expression assumes again the decoupling limit $M_A \gg M_Z$. For small Higgs masses, corrections at the $1/M$ level become important and are included in our numerical analysis. 
In writing~(\ref{eq:A}) we also assume that the non-holomorphic lepton-Higgs coupling $\lambda_\ell$ is proportional to the lepton Yukawa
\begin{equation}
\frac{m_S}{M} \lambda_\ell = \frac{m_S}{M} \varepsilon_\ell Y_\ell = \bar\varepsilon_\ell Y_\ell ~.
\end{equation}
As the NP contribution~(\ref{eq:A}) to the $B_s \to \mu^+ \mu^-$ amplitude grows with $\tan^3\beta$, the large $\tan\beta$ regime of the MSSM is strongly constrained by the experimental bound~(\ref{eq:Bsmm_LHC}).
However, as already stressed above, the non-holomorphic tree level Higgs-fermion couplings allow to generate NP contributions to B mixing already for moderate values of $\tan\beta$, where the bound from BR$(B_s \to \mu^+\mu^-)$ is considerably relaxed as long as the muon Yukawa coupling is not largely enhanced by the $\tan\beta$ resummation factors.

%%%%%%%%%%%%%%%%%%%%%%%%%%%%%%%%%%%%%%%%%%%%%%%%%%%%%%%%%%%%%%%%%%
\section{Numerical Analysis}
%%%%%%%%%%%%%%%%%%%%%%%%%%%%%%%%%%%%%%%%%%%%%%%%%%%%%%%%%%%%%%%%%%

To obtain the $B_q$ mixing amplitudes, we use 2-loop renormalization group running for the Wilson coefficients~\cite{Ciuchini:1997bw} and the hadronic matrix elements from~\cite{Becirevic:2001xt}.
We check compatibility of the model with various constraints.

(i) Vacuum stability: The higher dimensional operators in the superpotential can lead to a second minimum in the Higgs potential. Requiring that the electroweak minimum is stable, gives a lower bound on the charged Higgs mass for given values of the $\mu$ term and the SUSY breaking scale $m_S$~\cite{Blum:2009na,Altmannshofer:2011rm}.

(ii) Electroweak precision observables can constrain regions of parameter space where the dimension 5 superpotential operators lead to a very heavy SM like Higgs boson or a large splitting between the heavy Higgs bosons. We implement the S and T parameter following~\cite{Altmannshofer:2011rm}. Also the $Zbb$ coupling can be modified significantly by Higgs loops~\cite{Cline:2011mm,Degrassi:2010ne}. However the $Zbb$ constraint can be avoided if the Higgs-top coupling is suppressed by non-holomorphic corrections and therefore we do not include it in the numerical analysis.

(iii) Electric Dipole Moments (EDMs) can be induced both by phases of the higher dimensional operators in the K\"ahler potential and the superpotential. 
Experimentally accessible EDMs, like the EDMs of Thallium, Mercury or the neutron can be generated by the electron and quark EDMs and chromo EDMs (CEDMs), by CP violating 4 fermion operators~\cite{Barr:1991yx} as well as by the Weinberg 3 gluon operator~\cite{Weinberg:1989dx}.

The fermion (C)EDMs are generated at the 1-loop level by sparticle loops that are sensitive to the phase of the Higgs vev~\cite{Blum:2010by,Altmannshofer:2011rm} as well as to possible complex $1/M$ corrections to the sfermion mass matrices that are induced by modified Higgs-sfermion couplings after electroweak symmetry breaking. The latter corrections can however always be avoided if the parameters $\varepsilon_0$ and $\varepsilon_\ell$ are real.
At the 2-loop level, Barr-Zee diagrams contribute to the (C)EDMs~\cite{Barr:1990vd,Pilaftsis:2002fe,Buras:2010zm}. They are directly sensitive to both the phases in the non-holomorphic Higgs couplings and the scalar-pseudoscalar mixing in the Higgs sector.

The CP violating 4 fermion operators are induced by neutral Higgs exchange at tree level~\cite{Demir:2003js,Buras:2010zm} and, analogously to the 2-loop Barr-Zee contributions, they are sensitive to both the phases in the non-holomorphic Higgs couplings and the scalar-pseudoscalar mixing in the Higgs sector.

Finally, contributions to the Weinberg 3 gluon operator can be induced by 2-loop diagrams that are sensitive to the scalar-pseudoscalar mixing in the Higgs sector and in particular also to the phases in the non-holomorphic Higgs couplings~\cite{Trott:2010iz,Boyd:1990bx}. These contributions can be sizable, but they can be avoided to a large extent if the Higgs-top couplings are suppressed by non-holomorphic corrections. Keeping also in mind the large uncertainties in estimating the contribution of the 3 gluon operator to the neutron EDM, we do not include it in our numerical analysis.
Following this approach, we find that the most important contributions are typically 2-loop Barr-Zee contributions to the mercury EDM and 1-loop Higgsino-Wino-slepton contributions to the Thallium EDM in the regions of parameter space that we consider below. 

(iv) Constraints from direct Higgs searches at LEP, Tevatron and LHC are implemented using \verb|HiggsBounds|~\cite{Bechtle:2008jh} as well as the latest updates from Atlas and CMS~\cite{LHChiggs}. Generically direct SM Higgs searches do not lead to strong constraints, as the lightest Higgs boson is usually in the range $120 - 140$~GeV in the regions of parameter space considered below. The two heavier Higgs bosons can be much heavier and, due to the moderate values of $\tan\beta$, SUSY Higgs searches are also not constraining possible large effects in B mixing.

(v) Flavor observables: The main constraints come from $\Delta M_d$ and $\Delta M_s$ as well as the branching ratios of the decays $B \to X_s \gamma$ and $B_s \to \mu^+ \mu^-$. We also implement the constraints from $B \to \tau\nu$, $B \to D\tau\nu$ and $K \to \mu\nu$.

%%%%%%%%%%%%%%%%%%%%%%%%%%%%%%%%%%%%%%%%%%%%%%%%%%%%%%%%%%%%%%%
\begin{figure*}[t]  \centering
\includegraphics[width=0.48\textwidth]{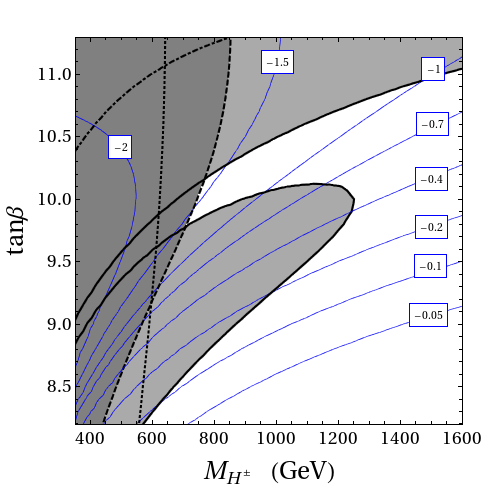} ~~~~~
\includegraphics[width=0.47\textwidth]{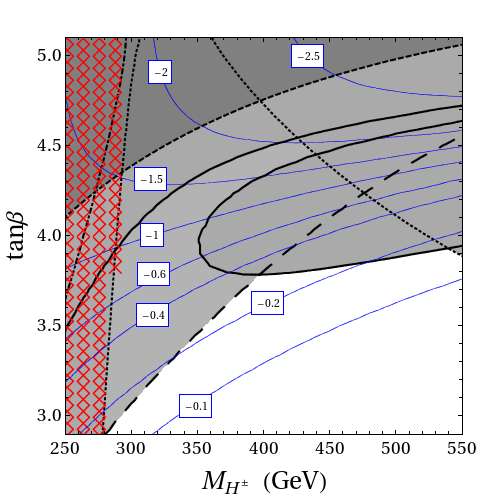}
\caption{
Possible values for the NP phase $\phi_s^{\rm NP}$ in the $M_{H^\pm}$ - $\tan\beta$ plane in the two example scenarios described in the text. The gray regions are excluded by $\Delta M_s$ (solid line), $\Delta M_d$ (long dashed line), BR$(B \to X_s \gamma)$ (dotted line), BR$(B_s \to \mu^+ \mu^-)$ (short dashed line) and the mercury EDM (dash-dotted line). In the red hatched region of the right plot, the electroweak vacuum is not absolutely stable.
}
\label{fig:MA_tanb}
\end{figure*}
%%%%%%%%%%%%%%%%%%%%%%%%%%%%%%%%%%%%%%%%%%%%%%%%%%%%%%%%%%%%%%%

In Fig.~\ref{fig:MA_tanb} we show in two representative scenarios the possible values of the NP phase in $B_s$ mixing in the $M_{H^+}$ - $\tan\beta$ plane together with the above mentioned constraints.
The plot on the left of Fig.~\ref{fig:MA_tanb} shows scenario I, where we chose a common sfermion mass of $\tilde m = 1$~TeV and trilinear couplings $A_t = \tilde m$, $A_b = A_\tau = 0$.\footnote{We remark that in a MFV framework, both the squark soft masses and the trilinear couplings can in general be expanded in powers of the flavor violating spurions $\mathcal{Y}_q = Y_q^\prime$. Higher order terms in the expansion can in principle induce additional flavor changing effects at the loop level. In the case of neutral meson mixing and $B_s \to \mu^+\mu^-$ however, the loop level effects induced by the higher order tems will be subdominant compared to the dominant tree level contributions from the modified Higgs sector. Among the considered flavor observables, only the loop induced $B \to X_s \gamma$ decay can be noticeably affected by possible higher order terms in the trilinear parameters. Considering such higher order terms would therefore add more flexibility in controlling the $B \to X_s \gamma$ constraint. As this constraint turns out to have only a small impact on our analysis, the higher order terms would not change any of our conclusions.} We assume the absence of the higher dimensional operators~(\ref{eq:super}) in the superpotential and set 
\begin{eqnarray}
\epsilon_0 &=& -0.8 ~,~ \epsilon_1 = 0 ~,~ \epsilon_2 = 0.3 ~, \nonumber \\
\epsilon_3 &=& 0~,~ \epsilon_4 = -0.5 -0.8i ~,~ \epsilon_\ell = 1 ~,\nonumber \\
\alpha &=& \omega = 0~~, \nonumber \\
\mu &=& 200\,\textnormal{GeV} ~,~ m_S = 1\,\textnormal{TeV} ~, \nonumber \\
M &=& 10\,\textnormal{TeV} ~.
\end{eqnarray}
Correspondingly, NP contribution to the $B$ mixing amplitudes are generated through the Wilson coefficient $C_4$ and therefore effects are much larger in $B_s$ than in $B_d$ mixing. We observe that the $B_s$ mixing phase can easily reach the $2\sigma$ bound given in~(\ref{eq:phi_s}), $\phi_s^{\rm NP} \gtrsim -0.4$, even for moderate values of $\tan\beta \simeq 10$ and very large Higgs masses of $M_{H^\pm} \gtrsim 1$~TeV.
In this region of parameter space, Higgs contributions to the $B_s \to \mu^+\mu^-$ decay and to $B \to X_s \gamma$ are well under control. The strongest EDM constraints in this scenario come from the mercury EDM but due to the large Higgs masses they turn out to be easily fulfilled. As the higher dimensional operators in the superpotential are not present, vacuum stability bounds as well as electroweak precision constraints are always fulfilled.

The plot on the right of Fig.~\ref{fig:MA_tanb} shows scenario II, where we allow both for higher dimensional operators in the superpotential and K\"ahler potential, but consider new sources of CP violation only in the superpotential. We chose third generation squark soft masses of $m_{\tilde q_L} = m_{\tilde t_R} = m_{\tilde b_R} = 700$~GeV, all remaining sfermion masses $\tilde m = 2 m_{\tilde q_L}$ and trilinear couplings $A_t = 2$~TeV, $A_b = A_\tau = 0$. In addition we set
\begin{eqnarray}
\epsilon_0 &=& 0 ~,~ \epsilon_1 = 0 ~,~ \epsilon_2 = -1.8 ~, \nonumber \\
\epsilon_3 &=& -1.6 ~,~ \epsilon_4 = 1.6 ~,~ \epsilon_\ell = 1~, \nonumber \\
\alpha &=& 1.5 - i ~,~ \omega = 1.5 ~, \nonumber \\
\mu &=& 150\,\textnormal{GeV} ~,~ m_S = 500\,\textnormal{GeV} ~, \nonumber \\
M &=& 5\,\textnormal{TeV} ~.
\end{eqnarray}
In this setup the NP phases in $B_s$ and $B_d$ mixing are induced by the Wilson coefficient $\tilde C_2$. Therefore they are of comparable size in both cases and typically well within the $2\sigma$ ranges of~(\ref{eq:phi_d}) and~(\ref{eq:phi_s}). We observe that sizable values for the $B_s$ mixing phase are possible even for small $\tan\beta \simeq 4 $ but require a rather light Higgs spectrum, which agrees with our expectation that $\tilde C_2$ can be important only for small Higgs masses.
Because of the small values of $\tan\beta$, both constraints from $B_s \to \mu^+ \mu^-$ and from the EDMs (that mainly come from the Thallium EDM) are well under control in the considered scenario. Concerning the bound from BR$(B \to X_s \gamma)$ we remark that for the light Higgs masses, there are sizable charged Higgs loop contributions to the $b \to s \gamma$ amplitude that are further enhanced by higher order $\tan\beta$ resummation factors. These contributions can be partly canceled either by $t_\beta m_S/M$ corrections to the couplings of the charged Higgs to the right-handed top quark, or by chargino-stop loops as long as stops are rather light, below 1~TeV. In contrast to scenario I, a very important constraint is now coming from $\Delta M_d$. Also bounds from vacuum stability start to be important. Contrary to the framework discussed in~\cite{Altmannshofer:2011rm} however, vacuum stability bounds can be compatible with a large $B_s$ mixing phase without the need of additional physics that stabilizes the electroweak vacuum.\footnote{We note that in the framework considered in~\cite{Altmannshofer:2011rm} a large negative $\mu$ term is required to generate sizable flavor changing neutral Higgs couplings at the loop level. In the scenario considered here however, small values for $\mu$ are possible that considerably soften the vacuum stability bounds.}

%%%%%%%%%%%%%%%%%%%%%%%%%%%%%%%%%%%%%%%%%%%%%%%%%%%%%%%%%%%%%%%
\begin{figure*}[t] \centering
\includegraphics[width=0.47\textwidth]{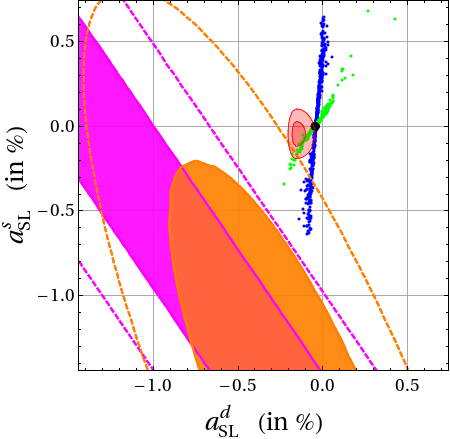} ~~~~~
\includegraphics[width=0.47\textwidth]{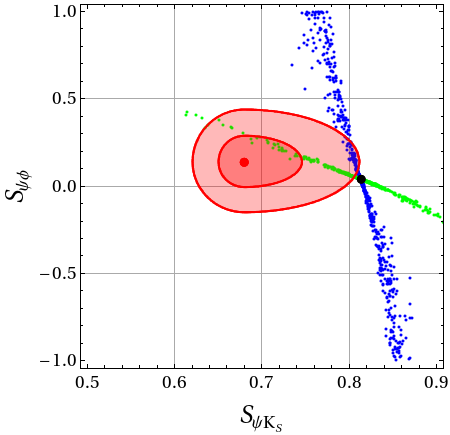}
\caption{
Correlation between the semi-leptonic asymmetries $a_{\rm SL}^d$ and $a_{\rm SL}^s$ (left) and the time-dependent CP asymmetries $S_{\psi K_S}$ and $S_{\psi\phi}$ (right). Blue (dark gray) points correspond to a scenario with new sources of CP violation in the Yukawa couplings, while green (light gray) points correspond to new CP phases in the Higgs sector. The red thin contours show the allowed regions using the bounds on $\phi_s^{\rm NP}$ and $\phi_d^{\rm NP}$ from Fig.~\ref{fig:phi_fit}. In the left plot, the magenta band (orange ellipse) shows the measurement of $A_{\rm SL}^b$ ($a_{\rm SL}^d$ and $a_{\rm SL}^s$) by D0.
}
\label{fig:aSL}
\end{figure*}
%%%%%%%%%%%%%%%%%%%%%%%%%%%%%%%%%%%%%%%%%%%%%%%%%%%%%%%%%%%%%%%

In Fig.~\ref{fig:aSL} we present the results of a parameter scan of the model. The left plot shows the correlation between the semi-leptonic asymmetries $a_{\rm SL}^d$ and $a_{\rm SL}^d$, while the right plot shows the correlation between the time dependent CP asymmetries $S_{\psi K_S}$ and $S_{\psi\phi}$. Blue (dark gray) points correspond to a scenario where new sources of CP violation are arising entirely from the modified Yukawa couplings and the higher dimensional operators in the superpotential~(\ref{eq:super}) are switched off completely. We consider mass scales as in scenario I above and allow $\tan\beta < 15$ as well as epsilon parameters $|\varepsilon_i| < 1$ with $\varepsilon_0$ and $\varepsilon_\ell$ real and arbitrary phases for the remaining $\varepsilon_i$.
As expected, in such a setup the NP effects in $B_s$ mixing are much larger than in $B_d$ mixing and the LHCb bounds on $\phi_s$ exclude a sizable NP phase in $B_d$ mixing.

Green (light gray) points correspond to a scenario where also the higher dimensional operators in the superpotential~(\ref{eq:super}) are considered. We fix mass scales as in scenario II above and allow the parameters $|\alpha|, |\omega| < 2$ with arbitrary phases and real epsilon parameters $|\varepsilon_i| < 2$ as well as $\tan\beta < 15$. In this setup, the NP contributions to CP violation in $B_d$ and $B_s$ mixing are comparable. Consequently, a sizable NP phase in $B_d$ mixing can be compatible with the LHCb constraints on the NP phase in $B_s$ mixing.  
Even though new CP phases come entirely from the superpotential operators, we stress that the presence of the K\"ahler potential operators is crucial. They allow for low $\tan\beta$ values and therefore the constraint from $B_s \to \mu^+\mu^-$ can be avoided.

We conclude that in order to generate sizable corrections to CP violation in $B_d$ mixing that are in agreement with the LHCb data on CP violation in $B_s$ mixing, both the higher dimensional operators in the superpotential (that modify the Higgs spectrum) as well as the higher dimensional operators in the K\"ahler potential (that lead to non-holomorphic Higgs-fermion couplings at tree level) are required.

%%%%%%%%%%%%%%%%%%%%%%%%%%%%%%%%%%%%%%%%%%%%%%%%%%%%%%%%%%%%%%%%%%
\section{Conclusions}
%%%%%%%%%%%%%%%%%%%%%%%%%%%%%%%%%%%%%%%%%%%%%%%%%%%%%%%%%%%%%%%%%%

Recent results from LHCb on the time-dependent CP asymmetries in $B_s \to \psi\phi$ and $B_s \to \psi f_0$ significantly restrict the allowed values for the $B_s$ mixing phase. Combining these results with measurements of CDF and D0 of the time-dependent CP asymmetry in $B_s \to \psi\phi$ as well as the measurements of the time-dependent CP asymmetry in $B_d \to \psi K_S$ at the B factories, we find the following 2$\sigma$ ranges for possible NP phases in $B_s$ and $B_d$ mixing: $-0.28 < \phi_d^{\rm NP} < 0$ and $-0.40 < \phi_s^{\rm NP} < 0.20$.
The preference for a negative NP phase in $B_d$ mixing is driven by tensions in fits of the Unitarity Triangle, while the NP phase in $B_s$ mixing is perfectly consistent with 0. Under the assumptions that neither the absorptive part of the $B_s$ mixing amplitude nor the SM tree-level $B_s \to \psi\phi$ decay amplitude are significantly affected by NP, the anomalous like sign dimuon charge asymmetry observed by D0 cannot be explained given the above bounds.\footnote{The possibility to reconcile the LHCb data on the time dependent CP asymmetries in $B_s \to \psi \phi$ and  $B_s \to \psi f_0$ and the D0 measurement of the like sign dimuon charge asymmetry with NP effects in $\Gamma_{12}$ has been analyzed very recently in \cite{Bobeth:2011st}.}

In view of these results we studied the possible impact of higher dimensional operators in the MSSM on B physics.
We considered dimension 5 operators both in the superpotential and in the K\"ahler potential assuming that they are generated at a scale of $M \simeq 5-10$~TeV. The 1/M suppressed operators in the superpotential have important impact on the Higgs spectrum. They can significantly enhance the tree level mass of the lightest Higgs boson and lead to a mass splitting between the two heavy neutral Higgs bosons. With complex coefficients they also lead to scalar-pseudoscalar mixing. The 1/M operators in the K\"ahler potential can induce non-holomorphic Higgs couplings and consequently flavor changing neutral Higgs couplings at the tree level. Assuming that the K\"ahler potential operators follow the minimal flavor violation ansatz, we find a flavor phenomenology that resembles to a large extent the 2 Higgs doublet model with MFV discussed in~\cite{Buras:2010mh,Buras:2010zm}.
In particular large values for $\tan\beta$ are not required to generate sizable corrections to B meson mixing from neutral Higgs exchange and therefore the strong constraint from BR$(B_s \to \mu^+ \mu^-)$ can be significantly relaxed. 

We find that both superpotential and K\"ahler potential operators are required to generate non-standard effects in the $B_d$ mixing phase that are in agreement with the current bounds on the $B_s$ mixing phase from LHCb. The corresponding region of parameter space is characterized by rather small values of $\tan\beta \simeq 5$ as well as low masses for the heavy Higgs bosons $M_{H_2}, M_{H_3} \simeq 300 - 400$~GeV.
We stress that in the class of models discussed in this work, a non-standard $B_d$ mixing phase does imply also non-standard effects in the $B_s$ mixing phase at a level that can be tested in the near future by LHCb.

\bigskip
%%%%%%%%%%%%%%%%%%%%%%%%%%%%%%%%%%%%%%%%%%%%%%%%%%%%%%%%%%%%%%%%%%
\paragraph{Acknowledgments:}
%%%%%%%%%%%%%%%%%%%%%%%%%%%%%%%%%%%%%%%%%%%%%%%%%%%%%%%%%%%%%%%%%%

We thank Michael Trott for useful comments.
Fermilab is operated by Fermi Research Alliance, LLC under Contract No. De-AC02-07CH11359 with the United States Department of Energy.

%%%%%%%%%%%%%%%%%%%%%%%%%%%%%%%%%%%%%%%%%%%%%%%%%%%%%%%%%%%%%%%%%%
\section{Appendix}
%%%%%%%%%%%%%%%%%%%%%%%%%%%%%%%%%%%%%%%%%%%%%%%%%%%%%%%%%%%%%%%%%%

Here we briefly present UV completions that lead to the higher dimensional operators in~(\ref{eq:super}) and (\ref{eq:kahler}). 
The dimension 5 superpotential operator in~(\ref{eq:super}) can be generated by integrating out a heavy singlet S with the following superpotential interactions~\cite{Dine:2007xi}
\begin{equation}
\mathcal{L} \supset \int d^2 \theta \left( \frac{1}{2} M S^2 + \sqrt{\omega} S H_u H_d \right) ~.
\end{equation}
Additional gauge interactions that are broken at a high scale $M$ can be effectively described by dimension 6 operators~\cite{Dine:2007xi,Carena:2009gx}.

One way to generate the K\"ahler potential operators in~(\ref{eq:kahler}) is to introduce two heavy $SU(2)$ doublets $\tilde H_u$, $\tilde H_d$ with hypercharge $+1$ and $-1$ that couple to fermions and mix with the MSSM Higgs doublets $H_u$, $H_d$~\cite{Antoniadis:2008es}.\footnote{To preserve gauge coupling unification in such a case, additional heavy matter fields should be present to complete SU(5) multiplets that include the heavy Higgs doublets.} Neglecting for simplicity gauge interactions, their K\"ahler potential and superpotential read
\begin{eqnarray}
\mathcal{L}&\supset& \int d^4 \theta \left( \tilde H_d^\dagger \tilde H_d + \tilde H_u^\dagger \tilde H_u \right)  \nonumber \\
&+& \int d^4 \theta \left( \tilde H_d^\dagger H_d + \tilde H_u^\dagger H_u  + {\rm h.c.} \right) \nonumber \\
&+& \int d^2 \theta \Big( M \tilde H_u \tilde H_d + \lambda_\ell \tilde H_d L E  \nonumber\\  \label{eq:UV}
& & ~ + \lambda_u \tilde H_u Q U + \lambda_d \tilde H_d Q D + {\rm h.c.} \Big) ~. 
\end{eqnarray}
Integrating out the heavy Higgs doublets generates the supersymmetric term in~(\ref{eq:kahler}) and SUSY breaking can be incorporated with the auxiliary spurion $Z$.

In addition to~(\ref{eq:kahler}), integrating out the heavy Higgses from~(\ref{eq:UV}) also generates $1/M$ suppressed terms in the superpotential
\begin{eqnarray}
\mathcal{L} &\supset& \frac{1}{M} \int d^2 \theta \Big( \lambda_u \lambda_d (QU)(QD) \nonumber \\
&& ~~~~~~ + \lambda_u \lambda_\ell (QU)(LE) + {\rm h.c.} \Big)~,
\end{eqnarray}
that in turn generate dimension 5 fermion -- sfermion interactions. Such interactions are however not relevant for the topics discussed in the present work.

%%%%%%%%%%%%%%%%%%%%%%%%%%%%%%%%%%%%%%%%%%%%%%%%%%%%%%%%%%%%%%%%%%

%%%%%%%%%%%%%%%%%%%%%%%%%%%%%%%%%%%%%%%%%%%%%%%%%%%%%%%%%%%%%%%%%%

\end{document}